\documentclass{article}

\def\IR{{ I\kern-.27em R}}

\usepackage[dvips]{graphicx}
\usepackage{amsmath,amssymb}
\usepackage{cite}

\title{Nonlocal symmetries of Riccati and Abel chains and their similarity reductions}
\author{M. S. Bruzon$^*$, M.L. Gandarias$^*$ and M. Senthilvelan$^{**}$}
\date{\small $^*$Departamento de Matem\'aticas, Universidad de C\'adiz,\\ PO.BOX
40,11510 Puerto Real, C\'adiz, Spain\\
$^{**}$Centre for Nonlinear Dynamics, Bharathidasan University,\\ Tiruchirappalli 620024, Tamil Nadu, India\\
}

\usepackage[dvips]{graphicx}
\usepackage{bm}
\begin{document}

\maketitle

\maketitle

\begin{abstract}
\noindent We study nonlocal symmetries and their similarity reductions of Riccati and Abel chains.  Our results show that all the equations in Riccati chain share the same  form of nonlocal symmetry.
The similarity reduced $N^{th}$ order ordinary differential equation (ODE), $N=2, 3,4,\cdots$, in this chain yields $(N-1)^{th}$ order ODE in the same chain.
 All the equations in the Abel chain also share the same form of nonlocal symmetry (which is different from the one that exist in Riccati chain) but the similarity reduced
 $N^{th}$ order ODE, $N=2, 3,4,\cdots$, in the Abel chain always ends at the $(N-1)^{th}$ order ODE in the Riccati chain.  We describe the method of finding general solution of all the equations
 that appear in these chains from the nonlocal symmetry.
\end{abstract}



\section{Introduction}
\subsection{Statement of the problem}
The first order Riccati equation,
\begin{equation}
u_x=f_0(x)+f_1(x)u+f_2(x)u^2,
\label{ric1}
\end{equation}
and its simplest nonlinear extension, namely Abel equation of first kind,
\begin{equation}
u_x=g_0(x)+g_1(x)u+g_2(x)u^2+g_3(x)u^3,
\label{abel1}
\end{equation}
where $f_i$'s and $g_j$'s, $i=0,1,2$ and $j=0,1,2,3$, are arbitrary functions of $x$ and subscript denotes derivative with respect to $x$, play a vital role in the theory of dynamical systems \cite{inc,dav,abl}.  Both the equations have been intensively studied by many authors and it has been shown that  both of them possess many interesting properties, see for example \cite{car1,car2} and references therein.

Interestingly, both the equations, (\ref{ric1}) and (\ref{abel1}) admit higher order integrable generalizations \cite{car2,eul,cha,gla1}. For example, the higher order Riccati equations/Riccati chain has the following form of differential equations, namely
\begin{eqnarray}
\label{eqq1}u_x+u^2&=&0,\\
\label{eqq2}u_{xx}+3uu_x+u^3&=&0,\\
\label{eqq3}u_{xxx}+4uu_{xx}+6u^2u_{x}+3u_{x}^2+u^4&=&0,\\
\label{eqq4} u_{xxxx}+5u u_{xxx}+10u_xu_{xx}+10u^2u_{xx}+15u u_x^2+10u^3u_x+u^5&=&0,\\ \nonumber
\end{eqnarray}
and so on.  The second ODE in
this family is the modified Emden equation \cite{vkc1} and the third
order ODE is one of the subcases of the Chazy equation \cite{vkc2}.

The higher order Abel equations/Abel chain has the following form of differential equations \cite{car2,gla1}, namely
\begin{eqnarray}
\label{eqqq1}u_{x}+u^3&=&0,\\
\label{eqqq2}u_{xx}+4u^2u_{x}+u^5&=&0,\\
\label{eqqq3}u_{xxx}+5u^2u_{xx}+8uu_{x}^2+9u^4u_{x}+u^7&=&0,\\
\label{eqqq4}u_{xxxx}+6u^2u_{xxx}+26u u_x u_{xx}+14u^4u_{x}+8u_x^3+44u^3u_x^2+16u^6u_x+u^9&=&0, \\ \nonumber
\end{eqnarray}
and so on.  The second order ODE is a generalized van-der Pol
oscillator equation \cite{vkc3} and the third order ODE is a subcase
of the Chazy equation \cite{vkc2}.

The properties exhibited by Riccati and Abel chains make them interesting from both physical and mathematical points of view \cite{car2}.  Very recently $n$-dimensional integrable generalizations of both the chains have also been proposed.

In this paper we study nonlocal symmetries and similarity reductions of Riccati and Abel chains.  The primary motivation for this study comes from the contemporary interest in studying symmetry, integrability and geometrical properties of both the chains. Even though the classical Lie point symmetries of both the chains have been studied the existence of nonlocal symmetries and their consequences for these two chains have not been analyzed in the literature.  The second reason for this study comes from the recent developments in exploring non-Lie point symmetries associated with the nonlinear differential equations
\cite{art,nucci,ada,barba00,barba11}.
The need for this
study came from the result that certain nonlinear integrable models do not admit Lie point symmetries.
To overcome this
demerit, during the past few years, several generalizations over the
classical Lie algorithm have been proposed.  Some of the algorithms that have been developed in the recent literature to derive integrals/general solution associated with the given ODE that lacks Lie point symmetries are $\lambda$-symmetries \cite{mur2},
telescopical symmetry \cite{puc}, hidden and non-local symmetries \cite{abr},
adjoint symmetry method \cite{blu2},
exponential vector fields \cite{olv},  generalized Lie symmetries \cite{sen}
 and so on.  In this paper we intend to study nonlocal symmetries associated with the given equation.
\subsection{Methodology and outcome}

To explore nonlocal symmetries of the given equation we consider the algorithm proposed by one of the present authors \cite{mltheo,gan2}.  In this method one essentially introduces an auxiliary ``covering''
system with auxiliary dependent variables.  A Lie symmetry of the
auxiliary system, acting on the space of independent
and dependent variables of the given ODE as well as the auxiliary
variables, yields a nonlocal symmetry of the given ODE if it does
not project to a point symmetry acting in its space of the
independent and dependent variables \cite{mltheo,gan2}.

Let the given second order nonlinear ODE be of the form
\begin{equation}
\Delta\bigg(x,u,\frac{du}{dx},\ldots,\frac{d^nu}{dx^n}\bigg)=0.
\label{eq1a}
\end{equation}
To derive nonlocal symmetries of (\ref{eq1a}), the author of
Ref.\cite{mltheo} has introduced an auxiliary nonlocal variable $v$
with the auxiliary system
\begin{equation}
\Delta\bigg(x,u,\frac{du}{dx},\ldots,\frac{d^nu}{dx^n}\bigg)=0,\;\;v_x=f(x,u).
\label{eq2}
\end{equation}

 Any Lie group of point transformations \cite{bluku1,olv,hydon}
\begin{equation}
{\bf v}=\xi(x,u,v)\frac{\partial}{\partial x}+\phi(x,u,v)
\frac{\partial}{\partial u}+\psi(x,u,v)\frac{\partial}{\partial
v},
\label{eq1b}
\end{equation}
admitted by (\ref{eq2}) yields a nonlocal symmetry of the given ODE
(\ref{eq1a}) if the infinitesimals $\xi$ or $\phi$ depend
explicitly on the new variable $v$, that is if the following
condition is satisfied \cite{blureid,ml1}
\begin{equation}
\label{cod}\xi_v^2+\phi_v^2 \neq 0.
\end{equation}
Because the local symmetries of (\ref{eq2}) are nonlocal symmetries of (\ref{eq1a}) this method
provides an algorithm to derive a class of nonlocal symmetries for the given equation.  These nonlocal symmetries can be profitably utilized to derive the general solution for the given
equation.

To begin with we consider Riccati chain.  We start our analysis with
second order Riccati equation (\ref{eqq2}).  By introducing a
nonlocal variable $v$ and rewriting it as a system of equations of
the form (\ref{eq2}) and extracting suitable point symmetries of
this auxiliary system of equations  we construct nonlocal symmetries
of Eq. (\ref{eqq2}). We then solve the characteristic equation
associated with the symmetries and obtain similarity variables which
in turn reduces the second order ODE into a first order nonlinear
ODE.  The first order ODE is nothing but the Riccati equation
(\ref{eqq1}).  We integrate this first order ODE and arrive at the
general solution of Eq. (\ref{eqq2}).  We then consider the third
order equation (\ref{eqq3}) and repeat the procedure.  While
reducing this third order ODE to a second order ODE we find that the
latter turns out to be the second order Riccati equation (\ref{eqq2}).
Since we have already derived the solution of this ODE in the
previous analysis we substitute the solution in the invariant and
integrate the resultant equation and obtain the general solution of
the third order equation (\ref{eqq3}).  Interestingly we observe
that this procedure is repetitive in every order.  For example, when
we carry out the symmetry reduction of fourth order ODE (\ref{eqq4})
we find that the similarity reduced ODE is nothing but the third
order equation (\ref{eqq3}) and the similarity reduced ODE from the
$N^{th}$ order Riccati equation is nothing but a $(N-1)^{th}$ order
Riccati equation.  The solution of any higher  order Riccati
equation can be constructed recursively from the known solution of
the previous ODE in this chain.

Interestingly when we extend this analysis to the Abel chain we find
that all the equations in this chain also admit the same form of
nonlocal symmetries (which is different from Riccati chain) but the
symmetry reduction of any equation in this chain always end at the
Riccati chain.  For example, the symmetry reduction of second order
Abel equation provides only the first order Riccati equation rather
than first order Abel equation.  Since we know the solution of this
first order equation in the Riccati chain we substitute it in the
invariant and integrate the resultant equation and arrive at the
general solution for the original equation.  Proceeding further we
find the symmetry reduction of third order Abel equation ends at
second order Riccati equation.  From the known general solution of
the latter we arrive at the general solution of the former .  In
fact, the result extends upto $N^{th}$ order ODE in the Abel chain.
To our knowledge the nonlocal symmetries and their similarity
reductions of these two chains are being presented for the first
time in the literature.

The plan of the paper is as follows.  In Sec. 2, we consider three equations in the Riccati chain, namely Riccati-$II,III$ and $IV$ and construct nonlocal symmetries for them.  We then construct similarity variable and reduce the order of these three equations to first, second and third order ODE respectively.  From the solution of these ODEs we derive the solution of the Riccati-$II,III$ and $IV$.   In Sec. 3, we consider three equations in the Abel chain, namely Abel-$II,III$ and $IV$ and study nonlocal symmetries, similarity reduction and general solution of them.  Here we show that the similarity reduction of all the Abel equations end at lower order Riccati equation.  We present our conclusion in Sec. 4.

\section{Nonlocal symmetries and similarity reduction of Riccati chain}
\subsection{ Riccati II}

Let us start our analysis with the second order Riccati equation
(\ref{eqq2}), namely
\begin{equation}
\hspace{5.8cm}u_{xx}+3u u_x+u^3=0.\nonumber \hspace{5.2cm}(\ref{eqq2})
\end{equation}
We introduce a nonlocal variable $v$ and rewrite Eq. (\ref{eqq2}) in the form
\begin{equation}\label{sents}\begin{array}{l}\displaystyle
u_{xx}+3u u_x+u^3=0\\
v_x=f(x,u),\end{array}\end{equation} where $f(x,u)$ is an arbitrary
function to be determined.  Any Lie group of point transformation
\begin{equation}\label{ge1}{\bf
v}=\xi(x,u,v)\frac{\partial}{\partial x}+\phi(x,u,v)
\frac{\partial}{\partial u}+\psi(x,u,v)\frac{\partial}{\partial
v},\end{equation}admitted
by (\ref{sents})
yields a nonlocal
symmetry of the ODE (\ref{eqq2}), if the infinitesimals $\xi$ and $\phi$  satisfy Eq.(\ref{cod}).

The invariance of the system (\ref{sents}) under a one parameter Lie group of
point transformations leads to the following set of determining equations, namely

\begin{eqnarray}
 \nonumber \xi_{uu}= 0,\;\;\psi_u-f\xi_u = 0,\;\;\phi_{uu}-f_u \xi_v-2 \xi_{ux}-2 f \xi_{uv}+6  u \xi_{u} &=& 0,\\\nonumber
    \psi_x+f \psi_v-f\xi_x-f^2\xi_v -f_x \xi-f_u \phi&=& 0, \\\nonumber 2 u^{3}
 \xi_x+2 f u^{3} \xi_v- \phi_u u^{3}+3 \phi u^{2}+3 \phi_x u+3 f \phi_v u
 +\phi_{xx}+f^{2} \phi_{vv}
 +2 f \phi_{vx}+f_x \phi_v&=& 0,\\ 3 u \xi_x-\xi_{xx}
 -f^{2} \xi_{vv}-2 f \xi_{vx}
 +3 f u \xi_{v}\label{ed} -f_x \xi_v+3 u^{3} \xi_{u}+f_u \phi_v+2
 \phi_{ux}+2 f \phi_{uv}+3 \phi &=& 0.
\label{sym1}
\end{eqnarray}

Solving the overdetermined system (\ref{sym1}) we obtain the infinitesimal generator of the form
\begin{equation}\label{gene2}
{\bf v}=c(x)e^{v}u \frac{\partial}{\partial u}+ c(x)e^{v}
\frac{\partial}{\partial v},
\end{equation}
with \begin{equation}\label{ff}f(x,u)= -u-\frac{c_x}{c},
\end{equation}
where $c(x)$ is an arbitrary function. We note here that (\ref{gene2}) is not the only solution set for the determining equations.

By using the  prolongation formula
\begin{equation}\label{gene3}
{\phi_x}=e^{v} (d_x u+d u_x+cuf)\frac{\partial}{\partial u_x}
\end{equation}
we find two functionally independent invariants are of the form
\begin{equation}
\label{inv1}\begin{array}{ll}z=x, &
\zeta=\displaystyle \frac{u_x}{u}+u.
\end{array}
\end{equation}
Differentiating  the second equation in (\ref{inv1}) and substituting the resultant expression in (\ref{eqq2}), we obtain a first order ODE of the form
\begin{equation}
\label{zh1}
\zeta_z+\zeta^2=0.
\end{equation}

 It follows that $\zeta=\displaystyle
\frac{1}{x+k_1}$ where $k_1$ is the integration constant.  Plugging the later expression in the second equation in (\ref{inv1}) we get
\begin{equation}
\frac{u_x}{u}+u-\frac{1}{x+k_1}=0.
\end{equation}
  This first order ODE can be integrated straightforwardly  to yield
 \begin{equation}\label{sol2}u= {{2\,\left(x+{
k_1}\right)}\over{x^{2}+2 { k_1}\,x-2\,
 { k_2}}}, \end{equation}
where $k_2$ is the second integration constant.  In this analysis we find that the similarity reduced first order ODE (\ref{zh1}) is nothing but the Riccati equation (\ref{eqq1}).
\subsection{Ricati III}

Now let us consider the next ODE (\ref{eqq3}) in the Riccati chain, namely
\begin{equation}
\hspace{4.5cm}u_{xxx}+4uu_{xx}+3u_x^2+6u^2u_x+u^4=0.\nonumber \hspace{4.0cm}(\ref{eqq3})
\end{equation}
To construct nonlocal symmetries of Eq. (\ref{eqq3}) we consider the system of equations
\begin{equation}\label{sents3}\begin{array}{l}\displaystyle
u_{xxx}+4uu_{xx}+3u_x^2+6u^2u_x+u^4=0\\
v_x=f(x,u).\end{array}\end{equation} 

The invariance of the system (\ref{sents3}) under a one parameter Lie group of
point transformations with infinitesimal generator  (\ref{ge1})
leads to a set of nine determining equations. We find that the infinitesimal generator (\ref{gene2}) with $f$ given in (\ref{ff}) again be a solution for these determining equations as well.  As a consequence the characteristic equation provides the same form of similarity variables (\ref{inv1}) for the Eq.(\ref{eqq3}).


The similarity reduced second order ODE turns out to be the second order Riccati equation (\ref{eqq2}), $\zeta_{xx}+3\,\zeta\,\zeta_{x}+\zeta^3=0,$ which has the general solution $(\ref{sol2})$.
Substituting $(\ref{sol2})$ into the second invariant that appears in (\ref{inv1}),
 \begin{equation}
\frac{u_x}{u}+u-{{2\,\left(x+{ k_1}\right)}\over{x^{2}+2 {
k_1}\,x-2\,
 { k_2}}} =0 ,\end{equation}\\
and solving the resultant ODE we obtain the solution of Riccati III in the form

  \begin{equation}\label{sol3}u(x)={{3\,\left(x^2+2\,{k_1}\,x-2\,{k_2}\right)}\over{x^3+3\,
 {k_1}\,x^2-6\,{k_2}\,x-3\,{k_3}}},\end{equation}\\
where $k_3$ is the third integration constant.  We note that the similarity reduced ODE of (\ref{eqq3}) is nothing but the second order Riccati equation $(4)$.

 \subsection{Riccati IV}
 For the sake of completeness we consider a next member in the Riccati chain, namely

\begin{equation}
\hspace{2.15cm}u_{xxxx}+5u u_{xxx}+10u_xu_{xx}+10u^2u_{xx}+15uu_x^2+10u^3u_x+u^5=0\nonumber\hspace{1.7cm}(\ref{eqq4})
\end{equation}
and consolidate the results. As we did in the previous two cases, we consider the system of equations,
\begin{equation}\label{sents4}\begin{array}{l}\displaystyle
 u_{xxxx}+5u u_{xxx}+10u_xu_{xx}+10u^2u_{xx}+15uu_x^2+10u^3u_x+u^5=0\\
v_x=f(x,u),\end{array}\end{equation}
 to derive nonlocal symmetries of (\ref{eqq4}).

The invariance of the system (\ref{sents4}) under a Lie group of
point transformation leads to a  set of fourteen equations. Here also we find that the infinitesimal generator (\ref{gene2}) with
$f$ given by (\ref{ff}) constitutes a solution for the determining equations.  Therefore, for the present case also we get same similarity variables (\ref{inv1}).  In terms
of these variables Eq. (\ref{eqq4}) can be brought to Riccati III (\ref{eqq3}), that is
\begin{equation}
\hspace{4.3cm}\zeta_{zzz}+4\zeta\zeta_{xx}+3 \zeta_x^2+6 \zeta^2 \zeta_x+\zeta^4=0.\nonumber\hspace{4.4cm}(\ref{eqq3})
\end{equation}
Since the solution of the latter is known, vide Eq. (\ref{sol3}), one can substitute it in (\ref{inv1}),
\begin{equation}
\frac{u_x}{u}+u-{{3\,\left(x^2+2\,{ k_1}\,x-2\,{ k_2}\right)}\over{x^3+3\,
 { k_1}\,x^2-6\,{ k_2}\,x-3\,{ k_3}}}=0
\end{equation}
and integrate to obtain the general solution of (\ref{eqq4}) in the form \begin{equation}
u(x,t)=-{{4\,x^3+12\,{ k_1}\,x^2-24\,{ k_2}\,x-12\,{ k_3}}\over{x
 ^4+4\,{ k_1}\,x^3-12\,{ k_2}\,x^2-12\,{ k_3}\,x+4\,
 { k_4}}},
\end{equation}
where $k_4$ is the fourth integration constant.

The above procedure can now be repeated to all higher order ODEs that present in the Riccati chain. The result reveals the fact that nonlocal symmetries of all the equations in this chain are same (vide Eqs. (\ref{gene2}) and (\ref{ff})) and the similarity reduced ODE always turns out to be the lower order equation in the same chain.
\section{Nonlocal symmetries and similarity reductions of Abel chain}

Now we focus our attention on the Abel chain and study the structure of nonlocal symmetries, similarity reductions and general solution admitted by this chain.

\subsection{Abel II}
To begin with let us consider the second order Abel equation (\ref{eqqq2}), namely
\begin{equation}
\hspace{5.65cm}u_{xx}+4u^2u_x+u^5=0.\nonumber \hspace{5.2cm}(\ref{eqqq2})\end{equation}
By introducing the nonlocal variable $v$ equation we have the auxilary system
\begin{equation}\label{abel2s}\begin{array}{l}\displaystyle
u_{xx}+4u^2u_x+u^5=0\\
v_x=f(x,u).\end{array}\end{equation}

The invariance of the system (\ref{abel2s}) under a Lie group of
point transformation leads to the following set of six equations, that is

\begin{eqnarray}
  \xi_{uu}= 0,\;\;\psi_u-f\xi_u = 0,\;\;\nonumber \phi_{uu}-f_u \xi_v-2 \xi_{ux}-2 f \xi_{uv}+8  u^2 \xi_{u} &=& 0, \\ \nonumber
    \psi_x-f\xi_x-f^2\xi_v -f_x \xi+f \psi_v-f_u \phi&=& 0, \\ \nonumber 2 u^{5}
 \xi_x+2 f u^{5} \xi_v- \phi_u u^{5}+5 \phi u^{4}+4 \phi_x u^2+4 f \phi_v u^2
 +\phi_{xx}+f^{2} \phi_{vv}
 +2 f \phi_{vx}+f_x \phi_v&=& 0,\\ \nonumber
 4 u^2 \xi_x-\xi_{xx}
 -f^{2} \xi_{vv}-2 f \xi_{vx}-f_x \xi_v+ 4 f u^2 \xi_{v}+3 u^{5} \xi_{u}+f_u \phi_v+2
 \phi_{ux}+2 f \phi_{uv}+8 \phi u &=& 0.\\
\label{new1}
\end{eqnarray}

Interestingly we find that the vector field
\begin{equation}\label{genea2}
{\bf v}=c(x)e^{v}u \frac{\partial}{\partial u}+ 2c(x)e^{v}
\frac{\partial}{\partial v},
\end{equation}
satisfies the determining equations (\ref{new1}) with the new form of $f(x,u)$ which is given by
 \begin{equation}\label{ff2}f(x,u)= -2u^2-\frac{c_x}{c}. \end{equation}

By using the  prolongation formula
\begin{equation}\label{genea3}
{\phi_x}=e^{v} (c_x u+c u_x+cuf)\frac{\partial}{\partial u_x}
\end{equation}
one can construct two functionally independent invariants of the form
\begin{equation}\label{inv2}\begin{array}{ll}z=x, &
\zeta=\displaystyle \frac{u_x}{u}+u^2.\end{array}\end{equation}
We observe that the second invariant is quadratic in $u$.  From (\ref{inv2}) we get $u_x=u(\zeta-u^2)$ and $u_{xx}=u\zeta_x+u_x\zeta-3u^2u_x$. In terms of these similarity variables, one can reduce the second order equation (\ref{eqqq2}) to a first order ODE, $\zeta_z+\zeta^2=0$,  which can be integrated to yield $\zeta=\displaystyle
\frac{1}{x+k_1}$, where $k_1$ is an integration constant.  Substituting this solution in the second expression in (\ref{inv2}),
 \begin{equation}
\frac{u_x}{u}+u^2-\frac{1}{x+k_1}=0,\end{equation}  and solving the resultant ODE we obtain the general solution of
$(\ref{eqqq2})$ in the form

 \begin{equation}u= {{\sqrt{3}\,\left(x+{k_1}\right)}\over{\sqrt{2\,x^{3}+6\,{
 k_1}\,x^{2}+6\,{k_1}^{2}\,x+3\,{k_2}}}},\end{equation}
where $k_2$ is the second integration constant.

We note that the similarity reduction of the  second order Abel ODE takes us only to the first order Riccati equation.

\subsection{Abel III}

Next we consider the third order ODE in the Abel chain (\ref{eqqq3}), namely
\begin{equation}
\hspace{4.2cm}u_{xxx}+5u^2u_{xx}+8uu_{x}^2+9u^4u_x+u^7=0\nonumber \hspace{3.6cm}(\ref{eqqq3})
\end{equation}
and explore the nonlocal symmetries of it. To do so we consider the
system
\begin{equation}\label{abelsis3}\begin{array}{l}\displaystyle
u_{xxx}+5u^2u_{xx}+8uu_x^2+9u^4u_x+u^7=0\\
v_x=f(x,u).\end{array}\end{equation}

The invariance of the system (\ref{abelsis3}) under a one parameter Lie group of
point transformations leads to a set of nine determining equations.
Solving this overdetermined system we obtain the infinitesimal generator
(\ref{genea2})
 with $f$ is given in $(\ref{ff2}).$
By solving the characteristic equation associated with the infinitesimal symmetries we obtain the same similarity variables given in (\ref{inv2}).  These  similarity variables reduce the third order ODE (\ref{eqqq3}) into second order Riccati equation, namely $\zeta_{xx}+3\,\zeta\,\zeta_{x}+\zeta^3=0,$  whose general solution is given by

\begin{equation}\label{zh2}\zeta= {{2\,x+{k_1}}\over{x^{2}+{k_1}\,x+{
 k_2}}}.\end{equation}
where $k_1$ ana $k_2$ are integration constants.
Substituting (\ref{zh2}) in the second invariant of (\ref{inv2}), we get
\begin{equation}
\frac{u_x}{u}+u^2 -{{2\,x+{k_1}}\over{x^{2}+{k_1}\,x+{
 k_2}}}=0.
\end{equation}
 This first order ODE can be integrated to yield

\begin{equation}
u= {{\sqrt{15}\,\left(x^{2}+{ k_1}\,x+{ k_2}%
 \right)}\over{\sqrt{6\,x^{5}+15\,{ k_1}\,x%
 ^{4}+\left(20\,{ k_2}+10\,{k_1}^{2}\right)\,x^{3}+30%
 \,{ k_1}\,{ k_2}\,x^{2}+30\,{ k_2}^{2}\,x+15\,{ k_3}}}},
\end{equation}
with  $4k_2-4k_1^2>0$,  where $k_3$ is the integration constant.

In this case also the similarity reduced Abel ODE ends at Riccati chain.
\subsection{Abel IV}
We conclude our analysis by considering the fourth order equation in the Abel chain (\ref{eqqq4}), namely

\begin{equation}
\hspace{1.5cm}u_{xxxx}+6u^2u_{xxx}+26uu_xu_{xx}+14u^4u_{xx}+8u_x^3+44u^3u_x^2+16u^6u_x+u^9=0.\nonumber \hspace{2.0cm}(\ref{eqqq4})
\end{equation}




We use the same procedure followed in the previous two equations and construct nonlocal symmetries of Eq. (\ref{eqqq4}).  We find the nonlocal symmetries of this equation also of the form $(\ref{genea2})$.  The similarity reduced ODE in this case turns out to be $\zeta_{xxx}+4\,\zeta\,\zeta_{xx}+3
\zeta_x^2+6\zeta^2 \zeta_x+\zeta^4=0,$ which is nothing but the third order Riccati equation.  From the solution of the latter we deduce the general solution to (\ref{eqqq4}) which is of the form




\begin{equation}u= {{\sqrt{105}\,\left(x^{3}+{ k_1}\,x^{2}+{ k_2}\,x+
 { k_3}\right)}\over{\sqrt{P(x)}
 }}, \end{equation}
where
$$\begin{array}{ll}P(x)=&30\,x^{7}+70\,
 { k_1}\,x^{6}+\left(84\,{ k_2}+42\,{ k_1}^{2}\right)
 \,x^{5}+\left(105\,{ k_3}+105\,{ k_1}\,{ k_2}\right)
 \,x^{4}\\&+\left(140\,{ k_1}\,{ k_3}+70\,{ k_2}^{2}\right)\,x^{3}
 +210\,{ k_2}\,{ k_3}\,x^{2}+210\,{ k_3}^{2}\,x+105\,k_4.\end{array}$$
where $k_1, k_2, k_3$ and $k_4$ are integration constants. Following the procedure given in this section one can construct the general solution of all higher order ODEs in this chain. In this work we have focussed our attention on exploring nonlocal symmetries associated with the equations in the Riccati and Abel chains and constructing their underlying solutions through similarity reduced lower order ODEs. One can also exploit the nonlocal symmetries to construct first integrals, see for example Ref.[29], associated with these equations. By integrating the first integrals one can obtain the general solution.

%

%
%
%

\section{Conclusions}

In this paper we have studied nonlocal symmetries of Riccati and Abel chains and studied similarity reductions coming out from them.  Our studies show that the entire Riccati chain of equations admit the same form of nonlocal symmetry (vide Eq. (\ref{gene2})).  As a consequence the similarity variables are also found to be same for all the equations in the chain.  The similarity reduced $N^{th}$ order ODE, $N=2, 3,4,\cdots$, ends at $(N-1)^{th}$ order ODE in the Riccati chain.  From the solution of the $(N-1)^{th}$ order ODE we derive the general solution for the $N^{th}$ order ODE.  We have observed that all the equations in the Abel chain also posses the same form of nonlocal symmetry (which is different from Riccati chain).  The similarity reduced $N^{th}$
order ODEs, $N=2, 3,4,\cdots$, in the Abel chain always ends at $(N-1)^{th}$ order equation Riccati chain.  Since we have already constructed the general solution of $(N-1)^{th}$ order equation in the Riccati chain we substitute this solution in the invariant and integrate it to arrive the general solution of the considered equation in Abel chain.  To our knowledge the nonlocal symmetries of these two chains and their associated similarity reductions are being reported for the first time in the literature. The procedure presented in this paper is simple, algorithmic and straightforward and moreover it is applicable to a class of nonlinear ODEs.

\newpage

\section{Acknowledgments} The support of DGICYT project
MTM2009-11875 and Junta de Andaluc\'ia group FQM-201 are gratefully acknowledged. The work of MS forms part of a Department of Science
and Technology, Government of India, sponsored the research project.

\label{lastpage}

\end{document}